\documentclass[sigconf]{acmart}
\AtBeginDocument{%
  }

\setcopyright{acmlicensed}
\copyrightyear{2018}
\acmYear{2018}
\acmDOI{XXXXXXX.XXXXXXX}
\acmConference[Conference acronym 'XX]{Make sure to enter the correct
  conference title from your rights confirmation email}{June 03--05,
  2018}{Woodstock, NY}
\acmISBN{978-1-4503-XXXX-X/2018/06}

\usepackage{booktabs}
\usepackage{multirow}
\usepackage{array}
\usepackage{makecell}
\usepackage{xcolor}        
\usepackage{colortbl}      
\usepackage{rotating}      
\usepackage{tcolorbox}
\usepackage{graphicx}
\usepackage{adjustbox}
\usepackage{mdframed}
\usepackage{listings}
\usepackage{xcolor}
\usepackage{xurl}

\lstdefinestyle{mypython}{
    language=Python,
    backgroundcolor=\color{gray!10},
    basicstyle=\ttfamily\small,
    keywordstyle=\color{blue},
    commentstyle=\color{green!50!black},
    stringstyle=\color{red},
    numbers=left,
    numberstyle=\tiny\color{gray},
    stepnumber=1,
    numbersep=8pt,
    showstringspaces=false,
    breaklines=true,
    frame=single
}

\newmdenv[
  backgroundcolor=gray!10,
  linecolor=black,
  linewidth=1pt,
  roundcorner=4pt,
  innertopmargin=6pt,
  innerbottommargin=8pt,
  innerleftmargin=8pt,
  innerrightmargin=8pt,
  skipabove=12pt,        
  skipbelow=12pt,        
  frametitlebackgroundcolor=black,
  frametitlefont=\bfseries\color{white},
  frametitlerule=false,
  frametitlealignment=\raggedright,
  frametitleaboveskip=6pt,
  frametitlebelowskip=6pt,
]{findingbox}

\definecolor{hm0} {RGB}{255,255,255}
\definecolor{hm1} {RGB}{240,246,243}
\definecolor{hm2} {RGB}{225,237,231}
\definecolor{hm3} {RGB}{210,228,220}
\definecolor{hm4} {RGB}{195,220,208}
\definecolor{hm5} {RGB}{180,211,196}
\definecolor{hm6} {RGB}{165,202,184}
\definecolor{hm7} {RGB}{150,193,172}
\definecolor{hm8} {RGB}{135,184,161}
\definecolor{hm9} {RGB}{120,175,149}
\definecolor{hm10}{RGB}{105,166,137}
\definecolor{hm11}{RGB}{ 90,157,125}
\definecolor{hm12}{RGB}{ 75,148,114}
\definecolor{hm13}{RGB}{ 60,139,102}
\definecolor{hm14}{RGB}{ 45,130, 90}
\definecolor{hm15}{RGB}{ 30,120, 78}
\definecolor{hm16}{RGB}{ 15,110, 86}

\definecolor{darkgreen}{rgb}{0., 0.6, 0.}
\definecolor{darkred}{rgb}{0.6, 0., 0.}

\newcommand{\approach}{TRAILS}
\newcommand{\RQone}{How does \approach~'s performance compare to current approaches?}
\newcommand{\RQtwo}{What is the cost of each step of \approach?}
\newcommand{\RQthree}{How stable is the decision of \approach~across reruns?}
\newcommand{\RQfour}{How many consistently and correctly labeled code do each approach label?}

\begin{document}

\title{Inferring Code Correctness from Specification}

\author{Florian Tambon}
\email{florian.tambon@uni.lu}
\affiliation{%
  \institution{University of Luxembourg}
  \city{Luxembourg}
  \country{Luxembourg}
}

\author{Mike Papadakis}
\email{michail.papadakis@uni.lu}
\affiliation{%
  \institution{University of Luxembourg}
  \city{Luxembourg}
  \country{Luxembourg}
}

\renewcommand{\shortauthors}{Tambon et al.}

\begin{abstract}

Large language models (LLMs) have become integral to modern software development, enabling automated code generation at scale. However, validating the correctness of LLM-generated code remains a critical and largely unsolved challenge. Existing approaches either rely on dynamic consensus across multiple code candidates - making them costly and difficult to scale - or on static reasoning that is susceptible to dynamic bugs and order bias. In this paper, we propose \approach~ (Targeted Reasoning Agreement via Inputs and Specifications), an approach that grounds LLM reasoning with concrete (input, output) pairs. \approach~ first generates diverse test inputs via category partitioning based on the specification, then executes them against the candidate code and prompts LLMs to assess whether the resulting input-output pairs conform to the specification - without ever reasoning over the code itself.  Scores are aggregated across inputs, to determines whether the program is likely correct. We evaluate \approach~ on two datasets, LiveCodeBench and CoCoClaNeL, across three LLMs (Qwen3Coder-30B, Devstral-Small-24B, and Olmo3.1-Instruct), comparing against HoarePrompt and a Zero-Shot Chain-of-Thought baseline. \approach~ improves Matthew Correlation Coefficient by up to 39\% relative to Zero-Shot COT and consistently outperforms HoarePrompt. Beyond accuracy, \approach~ demonstrates greater stability across seeded runs, reducing sensitivity to LLM non-determinism, and assigns correct labels to a larger set of unique code samples than competing approaches.
\end{abstract}



\keywords{Automated code generation, LLMs, Prompt engineering, Code correctness}


\maketitle

\section{Introduction}

Large language models (LLMs) have revolutionized software development by automating code generation at an unprecedented scale. Tools like GitHub Copilot, ChatGPT, and Claude now assist developers in generating code snippets, functions, and even entire classes. Recent surveys indicate that LLM adoption in code generation workflows has grown exponentially, with industry-wide deployment across startups and enterprises alike \cite{chen2024survey}. This capability has transformed the landscape of software engineering, shifting the focus from code writing to code review and validation. 

However, the widespread adoption of LLM-based code generation has revealed a critical gap: while these models excel at generating plausible-looking code, validating its correctness with regard to the original specification remains error-prone, complex, expensive and time-consuming. For experienced developers traditionnally debugging and testing is a already complex and time consuming task \cite{straubinger2023survey}, particularly when working with unfamiliar problem domains or complex specifications. With emergence of LLMs, developers starting using LLMs to offset this workload, resulting in critical step being skipped such as testing and verifying generated code \cite{fawzy2025vibe}. In parallel, a new category of (non-expert) developers with little to no experience emerged, empowered by the capability of LLMs. However, these users face added challenges compared to developers with actual experience and knowledge, and tend to overly on LLMs given their lack of experise as well as overtrust LLMs' answers \cite{feldman2024non, fawzy2025vibe, geng2025exploring}. The fundamental challenge underlying this gap is linked to the test oracle problem, with the absence of reliable oracle for LLMs generated code. In traditional software development, engineers invest significant effort in creating test suites before or alongside code development. While LLMs can generate test suite \cite{yang2024evaluation, chen2024chatunitest, hossain2025togll, schafer2023empirical}, they are prone to mistake, even more so when the underlying code is incorrect \cite{konstantinou2024llms, huang2024measuring}. This create a compound effect where incorrect code can be assessed as correct by incorrect test.  

Given this issue, recent researches have studied the possibility to verify generated code correctness. All these studies have the same key element, which is to establish whether an inconsistency exists between the code and specification, either dynamically or statically. One of the first approach, CodeT \cite{chen2022codet}, calculate a consensus set between multiple generated test cases and candidates test programs, selecting a likely correct code based on the highest consensus. In a similar fashion, Valentin et al. \cite{valentin2025incoherence} used multiple candidates programs over a large set of generated inputs to calculate incoherence between candidates solution in order to select a likely correct candidate or abstain. Both approach relies on concrete executions of the code, and so are reliant on the generated inputs. Moreover, the approach can be hard to scale, as it relies on multiple test candidates for the same specification. Finally, their underlying hypothesis is that LLMs' incorrect codes will tend to have a wide enough array of incorrect behavior, so that incoherence with the set of correct codes will be noticeable. On the other hand to this approach, HoarePrompt \cite{bouras2025hoareprompt} aims to leverage LLMs reasoning guided via Hoare's logic in order to statically verify the correctness of a code, by checking at each step the state of the program for inconsistency. While more flexible than previous approach, because of the abscence of concrete execution, it can fail to identify dynamic bugs. Moreover, because of the inherent single binary final decision, models can be prone to order bias \cite{shi2025judging} which can hamper stability of the approach.

To tackle this issue, we propose \approach~(\textbf{T}argeted \textbf{R}easoning \textbf{A}greement via \textbf{I}nputs and \textbf{S}pecifications), which uses concrete input execution results to ground the LLMs reasoning. \approach~ follows a dual scheme in which test inputs are first generated using category partitioning based on the specification to force diversity and trigger incorrect program behavior. Inputs are then validated by running them against the candidate code, to ensure that they conform to the program input format. Then, in the second step, test inputs are executed on the program to provide candidate output (program behavior). LLMs are then prompted to reason over the specification against the input-output pairs and characterize them as conforming or not. This step pushes the LLM to reason over the program behavior with respect to the specification and identify incorrect outputs. 

The advantage of our approach is that it forces the reasoning to occur exclusively on the specifications and not on the generated code, which may be incorrect and bias the LLMs responses \cite{konstantinou2024llms, huang2024measuring}. The results of multiple inputs are then aggregated, yielding the expectation that the program is correct if LLMs consider its outputs to be conforming to the specifications more often than not. As such, we leverage a threshold to determine our confidence on correctness, i.e., above a specific number of non-conforming input-output pairs the code is considered as likely incorrect.

In a way, our approach tackles the problem as a "low trust" system, where the only source of truth is emphasized to be the specification, and each subsequent step is treated with some guarding measures. To this end, we study the following Research Questions (RQs):

\begin{itemize}
    \item[\textbf{RQ1}] \RQone
    \item[\textbf{RQ2}] \RQtwo
    \item[\textbf{RQ3}] \RQthree
    \item[\textbf{RQ4}] \RQfour
\end{itemize}

We evaluate our approach on two different code datasets, LiveCodeBench \cite{jain2024livecodebench} and CoCoClaNeL \cite{bouras2025hoareprompt} on three different LLMs: Qwen3Coder-30B \cite{Qwen3-Coder}, Devstral-Small-24B \cite{Devstral-Small2} and Olmo3.1-Instruct \cite{Olmo-3.1}, all relatively recent and capable open-source models that can git on a consumer grade single GPU. We compare our approach against current state-of-the-art approaches on correctness inference of a single sample, HoarePrompt, as well as a baseline using Zero-shot Chain-of-Thoughts (COT) Reasoning \cite{wei2022chain}. Our results show that \approach~ improves performance in terms of Matthew Correlation Coefficient across the board and up to 39\% relatively to Zero-Shot COT and improves over HoarePrompt. Token consumptions mainly stems from additional repair on either intractable incorrect code (by design to ensure validity of input) or invalid inputs on CoCoClaNeL which expects standard input (stdin) for the code which models struggle with. \approach~ not only improves performance, but also is more correctly stable: over seeded runs, it preserves more the label assigned to a code than competing approaches. This makes our approach more robust to non-determinism of LLMs and limit the effect of the seed. Finally, we show that \approach~ consistently assign the correct labels to more unique code across datasets and models compared to baselines. There is nevertheless a sizeable overlap which can motivate further research on how to properly combined different approaches.

\section{Approach}

\subsection{Motivating example}

\begin{figure*}
    \centering
    \includegraphics[width=\textwidth]{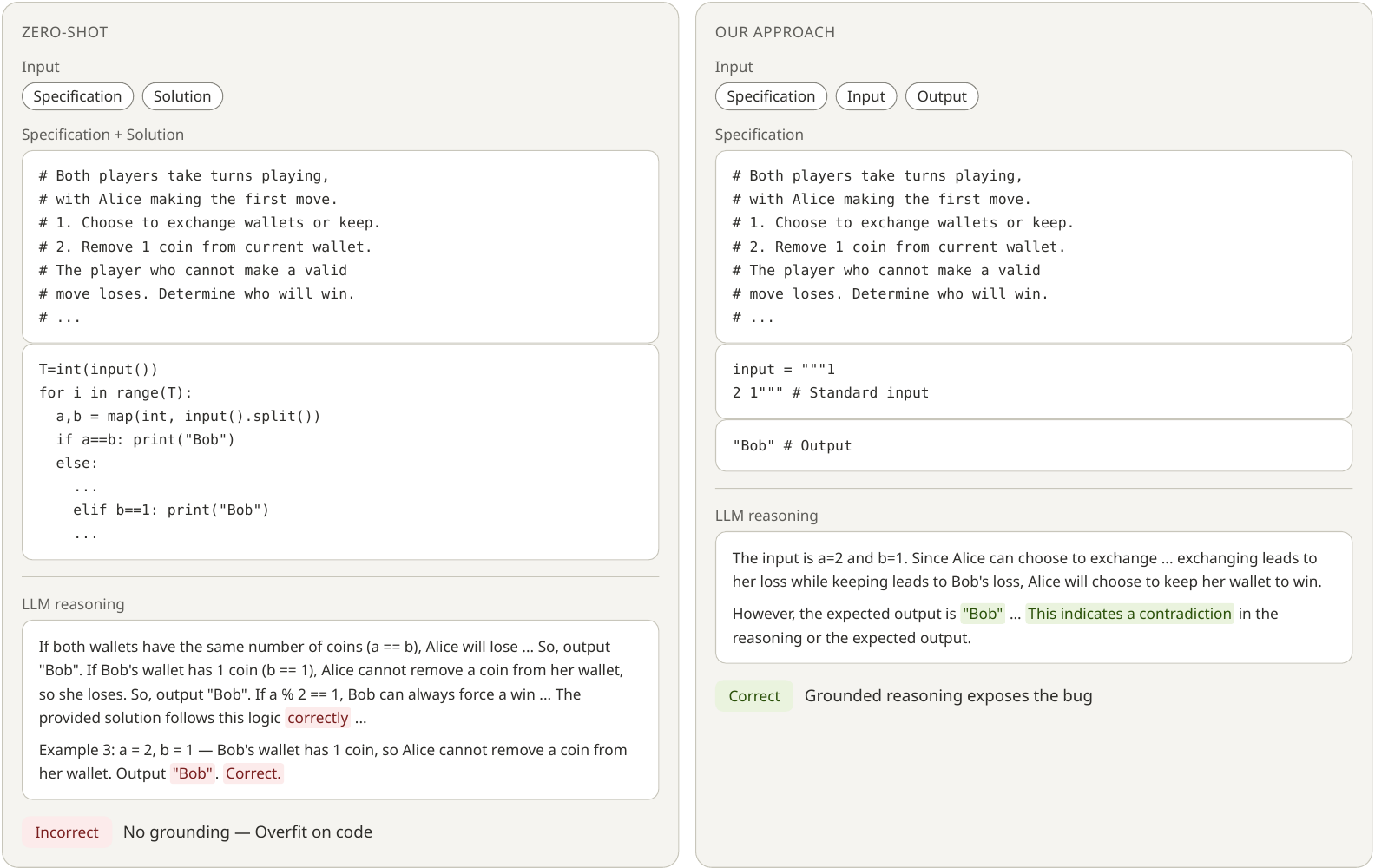}
    \caption{Motivation example. (Left) Zero-Shot reasoning: the model incorrectly validate the code logic. (Right) \approach~ correctly describe that the (input, output) pair can not be obtained with the given specification.}
    \label{fig:motiv}
\end{figure*}

Figure \ref{fig:motiv} illustrates how TRAILS work on an example from the CoCoClaNeL dataset. The task is to code a program that determines who between Bob and Alice will win a coin game. We compare our approach with traditional Zero-Shot COT. For traditional Zero-Shot COT, we provide the LLM with specification + the solution and ask the LLM to say whether the solution is correct or not with regard to the specification. The model incorrectly judge the code as correct. Even if it resorts to examples, it has no grounding output and so the LLM tends to follow on what the code does rather than reasoning over the specification as instructed to. In that case, it declares "Bob" to be the output for the input "a = 2, b = 1", even though the specification highlights that Alice goes first and so she should win. This behavior, LLM focusing on actual behavior of code rather than expected by specification, was already noted by previous research on test oracles \cite{konstantinou2024llms}. On the contrary, using our proposed \approach~, the LLM is provided with the specification, an input generated at an earlier stage and the executed output over the code solution. In that case, the LLM is asked to verify that the input-output pair is correct in relation to the specification. Since there is no code, it has to reason over the specification with the given input and, given the grounding output, highlight the contradiction between the expected behavior and actual one. In that case, it correctly exposes the bug.

\subsection{Problem formulation}

We consider the following problem. Given a specification in natural language (docstring, plain text etc.) and a code generated by LLMs which correctness is unknown, the goal is to infer the correctness of the code. Importantly, we consider that we have access to \textit{no} oracle, that is no available test case, no example nor user feedback to draw information from. This represent a general case of a user obtaining a code with little way of ensuring validity. As described in introduction, contrary to general benchmark such as HumanEval \cite{humanevalplus} which are composed of small code snippet where some test cases could easily be derived, practical usage of code LLMs often lead to code snippet where the user can not easily infer a proper oracle (especially, non-expert). While one could use possible test cases provided by the LLMs, studies \cite{huang2024measuring, schafer2023empirical} showed that this prone to error, especially when incorrect code is involved. Moreover, even with correct code, LLMs tend to generate incorrect or incomplete assertion, decreasing their accuracy, as noted in HoarePrompt study \cite{bouras2025hoareprompt}. Essentially, we set ourselves in a worst-case/low trust settings where the only source of trust is the specification. Moreover, contrary to current dynamic approaches which relies on the availability of \textit{N} code snippets, we only consider a \textit{single} code snippet, as asking for several code snippet is not easily scalable to larger code snippet.

\begin{figure*}
    \centering
    \includegraphics[width=0.9\textwidth]{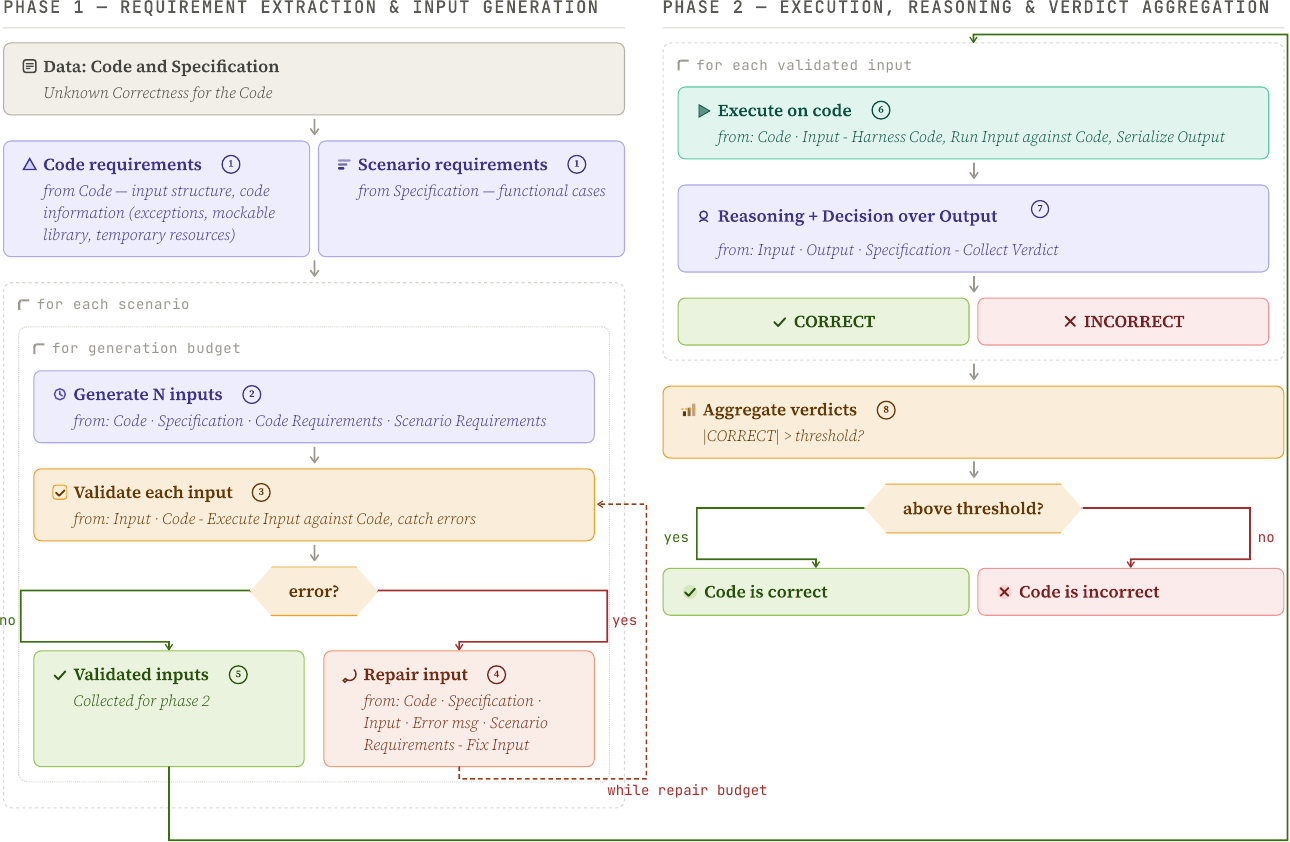}
    \caption{\approach ~overview}
    \label{fig:method}
\end{figure*}

\subsection{Overview of the approach}

An overview of the approach is represented on Figure \ref{fig:method}. The approach is divided into two phases: 1) input generations and 2) output verification. First, \approach~ aims to generate inputs that will be used to ground the reasoning of LLMs. To do so, it relies on categorical partioning to extract relevant behaviors and pre-conditions from the specification. The partitions are then used to obtain inputs encompassing different behavior pathes. Given inputs generated might not be valid, we adopt a simple repair/validation mechanism following existing studies on test generation \cite{konstantinou2025yate}. Essentially, the inputs are repaired under a given budget and deemed valid if they pass on the code under test, otherwise they are discarded. In the second stage, collected inputs are run on the concrete code and outputs are collected. Then, an LLM is prompted to verify the triplet (input, output, specification) via a binary decision. This verification is done for every inputs generated, resulting in an expected probability of the code being correct. Intuitively, we assume correct codes (i.e. code matching the specification) to lead to more verified outputs than incorrect codes.

\subsection{Framing the task inputs}

The aim of this step is to obtain an array of inputs that encapsulate the distribution of possible inputs the task is requiring. Asking directly an LLM to generate inputs might lead to edge cases being missed as the LLM will focus on common cases \cite{yang2024evaluation}. These edge cases might exactly be why the code of unknow correctness is not matching the specification. Applying mutations on top of generated inputs similarly to HumanEval+ \cite{humanevalplus} to increase diversity is not desirable in our case as it implies writting mutators for a large number of object types and use cases as well as verifying a high number of generated inputs with potentially broken conditions due to the mutators not being able to handle all possibilities. Given our aim is not develop a comprehensive set of inputs, but only to find possible discrepancy between code and specification, we rely on this much cheaper alternative. 

As such, we first prompt an LLM to map the possible behavior scenario based on the provided specification (1). The LLM is tasked with extracting a behavior, encompassing possible pre-conditions (e.g. "n should be strictly positive") as a free form text. Contrary to using logic formula, the free form text allows the LLM to represent more complex cases which might not be easily represented using such abstraction (e.g. handled exception in the code which should raise a dedicated error). In parallel, we also query an LLM to extract relevant properties of the code under test (1). This step is important to be able to properly execute the inputs down the line, providing the model with all relevant information. In particular, besides simple use case where inputs is either a list of arguments or a structure stdin, inputs could also require mockable dependency, temporary files, exception handling etc. For instance, when tackling a function sending GET request, it is imperative to be able to control the effect of the request to be able to cover all possibilites of the task, which is not doable via simple input injection. As such, the LLM is tasked with extracting properties such as the inputs structure, possible exceptions handling, mockable object or temporary resources needed. 

\subsection{Generation of valid input inputs}

Then, given an extraction partition behavior, the LLM is prompted to generate proper inputs for the code under test while accounting for the particular properties extracted (2). Then, each generated input is then validate by running it against the code under test (3). Any input leading to an unhandled crash of the code is deemed invalid. While this might exclude potentially valid and relevant inputs, this is nonetheless necessary following our low-trust philosophy: if a crash occurs, it might be because of a flawed code or an issue with the input, which is not trivial to decide. Nonetheless, given LLM might struggle on generating a valid input on some tasks, we add a simple repair mechanism with a given budget following current practice on test repair \cite{konstantinou2025yate}. Essentially, the LLM is prompted with the current input and the error raised and asked to correct the input, assuming a correct program (4). If no valid input can be generated in a given budget, even with repair, the partition is skipped and a new one is processed. If consecutive paritions lead to no valid input, an early stopping mechanism declare the code under test as invalid, preventing additional computation on a likely incorrect code. This steps leads, for each parition $p$ obtained, to a number $I_{tot, p}$ of inputs. To further optimize the process, a deduplication step (not shown in the graph) is ran by computing, within each partition, code coverage to remove inputs triggerring the same code lines. As such, one obtain a number of inputs per partition $I_{reduced, p} < I_{tot, p}$ (5).

\subsection{Output verifications}

The code is harnessed and the obtained inputs $I_{reduced, p}$ are ran against it to obtain the outputs $O_{reduced, p}$ (6). Each output is mapped onto an object that can be easily serializable for the verification step.

Given the triplets ($I_{reduced, p}$, $O_{reduced, p}$, specification), an LLM is prompted using COT to verify the output using binary labels outcome ('CORRECT'/'INCORRECT') and to explain its reasoning (7). While generating directly the oracle $O$ via reasoning is doable and could be potentially more informative, multiple studies highlight the difficulty for LLMs to generate proper oracle \cite{huang2024measuring, bouras2025hoareprompt, yang2024evaluation}, even more so when complex types are expected. Comparatively, using verification grounded on a concrete (input, output) pairs simplifies the process for the LLM. Intuitively, this also provide a grounding reference compared to plain Zero-Shot COT. Moreover, given we have multiple inputs per partition, we offset possible stochasticity issue due to binary decision. 

\subsection{Scoring over inputs}

In the end, the verdicts are aggregated we obtain a score per code under test (8). The obtained score represents the percentage of agreements across the inputs and is an approximation of the expected probability that the code under test is correct. Intuitively, the more agreement over inputs, the more likely the code is indeed correct. This aggregated score is compared to a threshold to decide whether the code is correct or not. This threshold might be obtained in two ways. The most straight forward is to use establish threshold over benchmarks. Since the score is in [0, 1], a higher threshold will lead to a more conservative approach (harder to accept a code as correct) at the cost of potentially discarding valid code. This is a trade-off that can be tuned by the user. An alternative is to tune the threshold on a calibration set that is provided based on the task to assess. While this might be harder to implement (it requires a dedicated calibration set), it gives more accuracy in the choice of threshold. In our experiment, we picked the threshold empirically, as experiments showed that one threshold gave best performance across models and datasets.

\section{Experimental setup}

\subsection{Datasets}

In our study, we use two datasets. LiveCodeBench \cite{jain2024livecodebench} is a dataset of competition programming extracted from different website, containing problem description with hidden test cases but no ground truth implementation. In this study, we use the v4 Lite versions, resulting in 119 tasks from half of 2024. 
To generate code implementations, we used a QwenCoder2.5-7B \cite{Qwen2.5-Coder} which was prompted to solve each task following the prompts given in the respective benchmarks. The model was prompted 10 times and obtained code where executed against the tests available in the benchmarks. For each task, we recovered a correct sample or an incorrect sample (or both) depending on the model performance. This model was not used in our study and is older than the LLMs used for evaluation to prevent bias. We also included CoCoClaNeL \cite{bouras2025hoareprompt} for comparison as it was introduced by HoarePrompt. It includes 163 tasks with 654 problem-solution pairs from Codeforces programming contests that took place in the first half of 2024, with annotated ground truth. Contrary to LiveCodeBench, this benchmark include human code of varying quality. Manually analyzing the obtained set, we noted 2 tasks with incomplete problem description which were removed from the set resulting in 161 tasks. For this dataset, we sampled two code (one correct/one incorrect) for each task, when possible (some tasks have no correct samples). 

\subsection{Models}

To evaluate our approach, we chose four models encompassing different architecture, training process and providers. As some baseline use their own reasoning as part of the process, to draw a fair comparison, we chose only models with no reasoning capability (or with capability turned off) to avoid influence of the reasoning over the results. We chose Qwen3Coder-30B \cite{Qwen3-Coder}, Devstral-Small-24B \cite{Devstral-Small2} and Olmo3.1-32B-Instruct \cite{Olmo-3.1}. The motivation to chose these particular models are multiple: they are all are mid-size open-source LLMs capable on fitting on a single consumer-grade GPU, they are from different providers (hence different architecture, training pipeline etc.) and do not have high performance on LiveCodeBench (and so CoCoClaNeL) contrary to fronteer closed-source models which allow to evaluate the usefulness of the approach rather than absolute models strength. Temperature was fixed to 0.5 for all models following HoarePrompt default settings. 

\subsection{Approaches}

As a baseline, we used both Zero-Shot Chain-of-Thoughts (COT) \cite{wei2022chain} and state-of-the-art HoarePrompt \cite{bouras2025hoareprompt}. For Zero-Shot COT, we use traditional prompt with the model needing to first reason (e.g. "Think step-by-step") before giving a verdict between "CORRECT" and "INCORRECT". The model is only given the code under test (of unknown correctness) and the specification, and must decided whether the model implement the specification. For HoarePrompt, we follow default parameters which use the whole pipeline (code + spec, no unrolling, precondition extraction, natural strongest postcondition and correctness classification) and also return "CORRECT" or "INCORRECT". We do not make any change to the prompts used. For \approach, we set the number of scenario to be generated to $s = 3$, the generation budget to $g = 3$ (i.e. at most 3 input/repair steps) and maximum deduplicated inputs per scenario to $|I_{reduced,p}| = 3$. In all cases, the temperature is set to 0.5 following HoarePrompt default configuration.



\begin{table*}[t]
  \centering
  \caption{Comparison of LLMs across two datasets using our approach and baselines.
Best result per dataset--metric block is in \textbf{bold}; second-best is \underline{underlined}. Results are averaged over three reruns. Tokens are given in thousands per task. Increase in performance is given relative to Zero-shot COT.}
  \label{tab:main_results}
  \Huge
  \renewcommand{\arraystretch}{1.15}
  \begin{adjustbox}{max width=\textwidth}
  \begin{tabular}{ll ccc ccc ccc |ccc |ccc }
    \toprule

    \multirow{2}{*}{\textbf{Dataset}} &
    \multirow{2}{*}{\textbf{Model}} &
    \multicolumn{9}{c}{\approach} &
    \multicolumn{3}{c}{HoarePrompt} &  
    \multicolumn{3}{c}{Zero-shot COT} \\
 
    \cmidrule(lr){3-11}
 
    & &
      \multicolumn{3}{c}{$\tau{=}0.6$} & \multicolumn{3}{c}{$\tau{=}0.7$} & \multicolumn{3}{c}{$\tau{=}0.8$} &
      & & & & & \\
 
    \cmidrule(lr){3-17}
    & &
      MCC ($\uparrow$) & P4 ($\uparrow$) & Tokens ($\downarrow$)&
      MCC ($\uparrow$) & P4 ($\uparrow$) & Tokens ($\downarrow$)&
      MCC ($\uparrow$) & P4 ($\uparrow$) & Tokens ($\downarrow$) &

      MCC ($\uparrow$) & P4 ($\uparrow$) & Tokens ($\downarrow$) &

      MCC ($\uparrow$) & P4 ($\uparrow$) & Tokens ($\downarrow$)\\
 
    \midrule
 
    \multirow{4}{*}{\rotatebox{90}{\textbf{LCB}}}
 
    & Qwen3-Coder
      & .633 \textcolor{darkgreen}{+ 3.43\%} & .776 \textcolor{darkred}{- 3.48\%} & 22,1k
      & \underline{.655} \textcolor{darkgreen}{+ 7.03\%} & \underline{.806} \textcolor{darkgreen}{+ 0.24\%} & 22,1k
      & \textbf{.661} \textcolor{darkgreen}{+ 8.01\%} & \textbf{.825} \textcolor{darkgreen}{+ 2.61\%} & 22,1k
      & .605 \textcolor{darkred}{- 1.14\%} & .798 \textcolor{darkred}{- 0.06\%} & 11,8k
      & .612 & .804 & 1,1k \\
    & Devstral-Small2
      & \textbf{.550} \textcolor{darkgreen}{+ 18.79\%} & .737 \textcolor{darkgreen}{+ 3.37\%} & 18,6k
      & \underline{.536} \textcolor{darkgreen}{+ 15.76\%} & \underline{.753} \textcolor{darkgreen}{+ 5.61\%} & 18,6k
      & .508 \textcolor{darkgreen}{+ 9.71\%} & \textbf{.754} \textcolor{darkgreen}{+ 5.75\%} & 18,6k
      & .357 \textcolor{darkred}{- 22.89\%} & .660 \textcolor{darkred}{- 7.43\%} & 16,7k
      & .463 & .713 & 1,2k \\
    & Olmo3.1-Instr
      & .579 \textcolor{darkgreen}{+ 24.78\%} & .773 \textcolor{darkgreen}{+ 6.92\%} & 21.9k
      & \underline{.601} \textcolor{darkgreen}{+ 29.52\%} & \underline{.793} \textcolor{darkgreen}{+ 9.68\%} & 21.9k
      & \textbf{.606} \textcolor{darkgreen}{+ 30.60\%}& \textbf{.803} \textcolor{darkgreen}{+ 11.07\%} & 21.9k
      & .355 \textcolor{darkred}{- 23.49\%} & .678 \textcolor{darkred}{- 6.22\%} & 11,3k
      & .464 & .723& 1,3k \\
 
    \midrule
 
    \multirow{4}{*}{\rotatebox{90}{\textbf{CoCo}}}
    & Qwen3-Coder
      & .211 \textcolor{darkgreen}{+ 13.44\%} & .500 \textcolor{darkred}{- 2.34\%} & 37,5k
      & \textbf{.259} \textcolor{darkgreen}{+ 39.24\%} & .591 \textcolor{darkgreen}{+ 15.43\%} & 37,5k
      & \underline{.246} \textcolor{darkgreen}{+ 32.26\%} & \textbf{.617} \textcolor{darkgreen}{+ 20.51\%} & 37,5k
      & .223 \textcolor{darkgreen}{+ 19.89\%} & \underline{.608} \textcolor{darkgreen}{+ 18.75\%} & 15,0k
      & .186 & .512 & 1,7k \\
    & Devstral-Small2
      & .214 \textcolor{darkgreen}{+ 0.94\%} & .521 \textcolor{darkred}{- 11.99\%} & 27,0k
      & \underline{.247} \textcolor{darkgreen}{+ 16.51\%} & .595 \textcolor{darkgreen}{+ 0.51\%} & 27,0k
      & \textbf{.261} \textcolor{darkgreen}{+ 23.11\%} & \textbf{.630} \textcolor{darkgreen}{+ 6.42\%} & 27,0k
      & .223 \textcolor{darkgreen}{+ 5.19\%} & \underline{.609} \textcolor{darkgreen}{+ 2.87\%} & 20,6k
      & .212 & .592 & 1,5k \\
    & Olmo3.1-Instr
      & .317 \textcolor{darkred}{- 4.51\%} & .598 \textcolor{darkgreen}{+ 1.18\%} & 30.3k
      & \underline{.411} \textcolor{darkgreen}{+ 23.80\%} & \underline{.684} \textcolor{darkgreen}{+ 15.74\%} & 30.3k
      & \textbf{.431} \textcolor{darkgreen}{+ 29.82\%} & \textbf{.711} \textcolor{darkgreen}{+ 20.30\%}& 30.3k
      & .269 \textcolor{darkred}{- 18.98\%} & .627 \textcolor{darkgreen}{+ 6.09\%} & 14,1k
      & .332 & .591 & 2,5k \\
 
    \bottomrule
  \end{tabular}
\end{adjustbox}
\end{table*}

\section{Results}

\subsection{RQ1: \RQone}

In our first research question, we compare the performance of \approach against other baselines. To track the performance, we use both Matthew Correlation Coefficients (MCC) \cite{MCC} similarly to HoarePrompt's study, as well as P4 metric \cite{sitarz2022extending}. P4 is a symmetric version of the F1 metric and considers also the whole confusion matrix as MCC does. Contrary to MCC, P4 put more emphasis on extreme cases (e.g. few false positive etc.) which is useful to contrast MCC value. Results are averaged over three reruns across datasets for each method/model. Results are given in Table \ref{tab:main_results}.

Overall, \approach~ improves performance over all models and datasets, yielding more correct classifications. We compared multiple thresholds and display thre three best in terms of metrics. $\tau = 0.8$ seems to be the best across the board with $\tau = 0.7$ being a close second. $\tau = 0.6$ have degraded performance compared to Zero-shot especially in terms of P4. At similar model/dataset, relative performance gain is more emphasize on CoCoClaNeL compared to the Zero-shot COT. This might simply be explained by LiveCodeBench's task being more simple, and so there is less of an improvement to be made, the ceiling of performance being the performance of the model itself. We note the degraded performance of HoarePrompt on LiveCodeBench. We explain this decrease again by the "simpler" tasks of LiveCodeBench: in that case, the added logical reasoning might artificially complexify the reasoning trace compared to simple CoT, degrading performance. On the contrary, CoCoClaNeL involves coding competition code, written by human and so more complex specification as well as harder to process code. For instance, such code snippets often contains meaningless variable identifier ("a = ..., ii = ...") which were shown to degrade models' performance in reasoning \cite{wang2023does}. In that instance, \approach~'s triplet (input, output, specification) can help mitigate the reasoning degradation by grounding it with concrete inputs. Similarly, HoarePrompt's logic based reasoning improves over basic CoT, however at a smaller rate than \approach~ with accounting for $\tau=0.8$.

\begin{findingbox}[frametitle={Finding RQ1}]
\approach~improves performance in most of the cases compared to baselines, up to $39\%$ in terms of MCC and $20\%$ for P4. The improvement is particularly stricking on the more complex CoCoClaNeL dataset, thanks to the grounding (input, output) pairs that help stirring the reasoning. 
\end{findingbox}

\subsection{RQ2: \RQtwo}

In this research question we study the cost of our approach. We give in Table \ref{tab:main_results} the aggregated cost per task for each approach. Compared to similar approach, \approach~ exhibits a cost overhead in terms of tokens, roughly twice HoarePrompt tokens. We examine the distribution of the tokens cost: in practice, if the phase 2 (output verdict) is on average the most expensive part, it is proportional to the number of inputs generated. Relative to the number of inputs, this part do not contribute much to the overhead. The part that significantly contributes the performance cost is the repair part in phase 1: in practice, especially when the code under test is incorrect with a crashing mistake, the repair can incur half of the total cost. We compared using a MannWhitney test \cite{mannwhitneyu} the distribution of tokens for correct and incorrect code samples. For LiveCodeBench, in all case, the p-value was significant at threshold $0.05$ with small to medium size effect. As such, correct code assessment would cost a more similar token cost to HoarePrompt's, but the incorrect code assessment led to additional cost. This is nonetheless necessary to ensure the proper validity of the inputs with regard to the task. In the case of CoCoClaNeL, we did not observe statistically significant difference: both correct/incorrect code required repair. Analyzing the log, we note that most of the repair occurs because of mistakes in the input format with the model do not generate the correct amount of input arguments for the task. All CoCoClaNeL tasks requires to inject inputs via standard inputs, with particular constraints. As such, models might not have been widely trained on such input type and so would need guidance in that step, compared to baselines that do not generate any inputs. Finally, we note that for some models such as Devstral, the LLM struggle to follow the required formatting (inputs, in python block etc.), which further increase the cost. This could, nonetheless, be mitigated via dedicated fine-tuning of the model or optimisation of the given prompt to a particular model using dedicated prompt optimization technique \cite{agrawal2025gepa}. 

\begin{findingbox}[frametitle={Finding RQ2}]
\approach~tokens cost mainly stem from the repair steps incurred both by the necessity to repair inputs even if correct in the case of an incorrect code, as well as particular input format of CoCoClaNeL dataset (standard input). While higher than baselines, \approach~tokens cost remain of the same order of mangitude than HoarePrompt and could be further optimize via fine-tuning.
\end{findingbox}

\subsection{RQ3: \RQthree}

\begin{table*}[t]
  \centering
  \caption{Comparison of verdict stability across two datasets using our approach and baselines. Best result per dataset--metric block is in \textbf{bold}, second best is \underline{underlined}. 'Correct' is the number of correct code under test for which all runs of one approach labeled it as such (resp 'Incorrect'). We also give the percentage of tasks all labeled 'Correct' (resp. 'Incorrect') which are reachable (i.e. at least one run that labeled the task with this label). For \approach, we use $\tau = 0.8$ which yielded the best performance in RQ1.}
  \label{tab:stable}
  \begin{tabular}{ll cc | cc | cc }
    \toprule

    \multirow{2}{*}{\textbf{Dataset}} &
    \multirow{2}{*}{\textbf{Model}} &
    \multicolumn{2}{c}{\approach} &
    \multicolumn{2}{c}{HoarePrompt} &  
    \multicolumn{2}{c}{Zero-shot COT} \\
 
    \cmidrule(lr){3-8}
    & &
      Correct & Incorrect &
      Correct & Incorrect &
      Correct & Incorrect \\

    \midrule
 
    \multirow{4}{*}{\rotatebox{90}{\textbf{LCB}}}
 
    & Qwen3-Coder
      &
      57 (\underline{60.00\%}) & 53 (\textbf{98.15\%}) &
      55 (57.89\%) & 48 (\underline{88.89\%}) &
      49 (\textbf{63.63\%}) & 59 (81.94\%)\\
    & Devstral-Small2
      &
      44 (45.36\%) & 47 (\textbf{90.38\%})&
      54 (\underline{46.55\%}) & 26 (78.79\%)&
      57 (\textbf{49.57\%}) & 29 (\underline{85.29\%})\\
    & Olmo3.1-Instr
      &
      49 (\textbf{56.98\%}) & 54 (\textbf{85.71\%}) &
      43 (45.74\%) & 43 (\underline{78.18\%}) &
      38 (\underline{48.72\%}) & 54 (76.06\%)\\

    \midrule
 
    \multirow{4}{*}{\rotatebox{90}{\textbf{CoCo}}}
    & Qwen3-Coder
      &
      81 (\textbf{35.37\%}) & 59 (\textbf{74.68\%})&
      56 (30.43\%) & 83 (\underline{66.93\%})&
      33 (\underline{32.35\%})& 115 (55.82\%)\\
    & Devstral-Small2
      &
      67 (\underline{31.02\%}) & 68 (\textbf{73.91\%}) &
      70 (\textbf{31.39\%}) & 62 (\underline{72.94\%}) &
      48 (29.09\%) & 87 (60.84\%) \\
    & Olmo3.1-Instr
      &
      100 (\textbf{47.17\%}) & 77 (\textbf{80.21\%}) &
      51 (28.18\%) & 88 (\underline{69.29\%}) &
      37 (\underline{36.63\%}) & 129 (62.32\%) \\

    \bottomrule
  \end{tabular}%
 
\end{table*}

In the third research question, we focus on stability of the decision. On top of absolute performance, decision made by an approach should be correctly consistent across runs. That is, reruning with the same code/specification should not yield a different labels, otherwise it means the approach is sensitive to non-determinism of the model. To verify it, for each approach, we collected tasks across reruns for which: a) all runs labeled the task as "correct" and the code under test was correct, b) all runs labeled the task as "incorrect" and the code under test was incorrect. We then compute the ratio using this number of tasks with all reachable tasks (i.e. task for which one run at least assigned a given label), giving us a score of how "correctly" stable the approach is. A perfect classifier would be consistent across runs while predicting correctly. The naive classifier (always correct/incorrect) would end up with a low score on the naive class given all tasks would be reachable even if it is very stable. Similarly for a random classifier. As such, this score informs on the stability while accounting for possible bias. We give the results are given in Table \ref{tab:stable}.

From the table, we note that \approach~ is overall the most stable while correctly labeling the code under test, illustrated with the among the highest score across the board compared to the baselines. It is especially more correctly stable when tackling incorrect code with a score of 70\%+, meaning it generally consistently assign the correct label to invalid code and do not assign it to correct code. In comparison, Zero-Shot COT tend to be somewhat correctly stable when dealing with correct code, yet it tends to over assign "incorrect" label to code under test across runs, illustrated by the lower score, 10 to 20 percentage point lower than our approach. It yields similar results when labeling correct code, with a decrease in performance on LiveCodeBench/Olmo (8 percentage point difference). HoarePrompt exhibits a similar patterns: on some models it tends to exhibit a bias pronounced bias towards the "incorrect" labels. For "correct" labels, HoarePrompt is similar to other approaches. 

Intuitively, our \approach~'s grounding via diverse inputs make them less sensitive to variation due to non-determinism. It's less the case for Zero-shot COT or HoarePrompt reasoning which is drastically influenced by the reasoning path taken by the LLM, given there is no grounding inputs, and so more non-determinism \cite{ahn2024largelanguagemodelsmathematical, potamitis2025reasonbench}. From the user point of view, while Zero-COT or HoarePrompt can yield adequate performance for a lower cost, it comes with at the cost of a lower reliability on a particular individual task. On the contrary \approach~ allows for better stability. The results particularly indicate that this phenomenon is more prevalent on incorrect code, which might lead to detrimental result for the user (determining a code as correct when it's incorrect) in case of an unlucky seed.


\begin{findingbox}[frametitle={Finding RQ3}]
\approach~ shows better correct stability across models/datasets, labeling more consistenly the task with the correct label thanks to the grounding inputs. On the contrary, Zero-shot COT and HoarePrompt presents lower stability on tasks due to non-determinism, yielding to a lower decision trust on particular inputs. In particular, the stability is quite lower on incorrect code, leading to a higher possibility of deeming the code correct because of an unlucky seed.
\end{findingbox}

\subsection{RQ4: \RQfour}

\begin{table}[ht]
\centering
\caption{Number of consistently and correctly labeled code by each approach or intersection of approaches across datasets and models. \textbf{T} = \approach, \textbf{H} = HoarePrompt, \textbf{C} = Zero-Shot COT}
 \label{tab:match}
\vspace{6pt}
 \begin{adjustbox}{max width=\columnwidth}
\begin{tabular}{r l l | ccc | cccc }
  \toprule
  & \textbf{Model} & &
    \scriptsize only T &
    \scriptsize only H &
    \scriptsize only C &
    \scriptsize T$\cap$H &
    \scriptsize T$\cap$C &
    \scriptsize H$\cap$C &
    \scriptsize T$\cap$H$\cap$C \\
  \midrule
 
  \multirow{6}{*}{\rotatebox[origin=c]{90}{\small\textbf{LiveCodeBench}}}
  & \multirow{2}{*}{Qwen3-Coder}
    & \scriptsize Correct   & 11  & 2 & 0 & 6 & 2 & 9 & 38 \\
  & & \scriptsize Incorrect & 7 &  2 & 6 & 2 & 9 &  9 &  35 \\
  \cmidrule(l){2-10}
  & \multirow{2}{*}{Devstral-Small}
    & \scriptsize Correct   & 7 & 3 & 4 & 2 & 4 & 18 & 31 \\
  & & \scriptsize Incorrect & 20 & 3 & 3 &  7 & 10 & 6 & 10 \\
  \cmidrule(l){2-10}
  & \multirow{2}{*}{Olmo3.1}
    & \scriptsize Correct   & 16 &  4 & 0 & 5 & 4 & 10 & 24 \\
  & & \scriptsize Incorrect & 9 & 2 & 3 & 2 & 12 &  8 &  31 \\
  \midrule
 
  \multirow{6}{*}{\rotatebox[origin=c]{90}{\small\textbf{CoCoClaNeL}}}
  & \multirow{2}{*}{Qwen3-Coder}
    & \scriptsize Correct   & 48 & 12 & 1 & 15 & 3 & 14 & 15 \\
  & & \scriptsize Incorrect & 14 &  5 & 27 &  1 & 11 &  44 & 33 \\
  \cmidrule(l){2-10}
  & \multirow{2}{*}{Devstral-small}
    & \scriptsize Correct   & 25 & 15 & 5 & 16 & 4 & 17 & 22 \\
  & & \scriptsize Incorrect & 22 & 8 & 24 & 7 & 16 & 24 & 23 \\
  \cmidrule(l){2-10}
  & \multirow{2}{*}{Olmo3.1}
    & \scriptsize Correct  & 57 & 9 & 5 & 14 & 4 & 3 & 25  \\
  & & \scriptsize Incorrect & 8 & 2 & 21 & 2 & 24 & 41 & 43  \\
  \bottomrule
\end{tabular}
\end{adjustbox}
\end{table}

Finally, in the last research questions, we study differences in tasks always well labeled (i.e. always correct for correct code and always incorrect for incorrect code) across approaches. The goal is to see the number of tasks each approach is better at handling as well as matching behavior between approaches. Results are given in Table \ref{tab:match}. 

As one can observe, there is a sizable overlap between the approaches ($T\cap H \cap C$) indicating tasks for which, for a given models, all approach consistenly assign the correct labels. This behavior is, proportionally, more prominent on LiveCodeBench given the lower number of codes ($\sim$70 correct/incorrect codes) than on CoCoClaNeL ($\sim$150 correct/incorrect codes), which is likely because LiveCodeBench's tasks are on average simpler and the code was produced by an LLM. \approach~generally exposes more unique behavior on LiveCodeBench for both correct/incorrect labels and on CoCoClaNeL only for correct labels, highlighted by the higher number of tasks that are only labeled consistently by our approach. Such behavior are generally because the code contains long conditions or complex processing. In that case, the model is not burdened by reasoning over the code process itself. We give an example of such task in Listing \ref{listing_error}. In that example, both Zero-Shot COT and HoarePrompt will struggles processing the code, hampered by reasoning on the concrete code.

\begin{lstlisting}[float=tbp, style=mypython, caption={Correct code consistently misidentified by baselines. (LiveCodeBench - 9ccb)}, label={listing_error}]
from typing import *
class Solution:
    def removeAlmostEqualCharacters(self, word):
        result = 0
        for i in range(len(word)-1):
            if (i+1)+result >= len(word):
                break
            if abs(ord(word[(i+1)+result])-ord(word[i+result])) <= 1:
                result += 1
        return result

# HoarePrompt typical output
...
{ result is the maximum number of consecutive characters in word where each adjacent pair has ASCII values differing by at most 1, starting from some initial position. } # Ignore first stop condition
return result
{ the state is unknown }
...

# Zero-shot COT typical output
...
Let me trace through a simple example:
word = "abc"
- i=0: check if abs(ord('b')-ord('a')) <= 1 -> abs(98-97) = 1 <= 1, so we increment result to 1
- i=1: check if abs(ord('c')-ord('b')) <= 1 ->  abs(99-98) = 1 <= 1, so we increment result to 2 # Focus on the second condition, when the first one would have been triggered already
...
\end{lstlisting}

On the other hand of the spectrum, Zero-Shot COT highlights very few unique behavior in terms of correct code, but do so for incorrect code. As seen in previous research question, this might be partially explain by the Zero-shot COT being inclined to label code as incorrect. Zero-Shot COT and HoarePrompt tends to be more similar in consistently labeling correct code (higher number of tasks in the intersections $H \cap C$) in LiveCodeBench and in consistenly labeling incorrect code in CoCoClaNeL. In practice, this means there is complementarity in two or three combinations of approaches, meaning they could potentially be combined to further improve the classification performance.

\begin{findingbox}[frametitle={Finding RQ4}]
\approach~ consistently labels more unique code for both correct and incorrect labels, identifying more unique behavior compared to the baselines. That is, \approach~ generally correctly classify code that other approaches struggles with. However, there are sizeable overlap between all approaches, highlighting complementary between approaches. Zero-Shot COT and HoarePrompt, being pure reasoning, tends to label consistently similar tasks compared to \approach.
\end{findingbox}

\section{Approach's limitations}

Despite its advantages, \approach~ can not classify correctly all code under test. We thus dwelled into the potential limitations of the approach. To do so, we manually analyzed code under test for which, on a given dataset, at least two models labeled consistently the code under test with the opposite label using \approach. We identified the following main issues:

\subsubsection*{Valid inputs causing crash on incorrect code}

We give an example of such case in Listing \ref{listing1}. In that code, there is a bug triggered when x > y with the increment operation causing a recursion issue. The rest of the code is working correctly (x == y and x < y). Despite this being an identified scenario by \approach~, given inputs in this scenario will always cause a failure, such inputs are not consider for verification. In that case, the LLM always assert that correct scenarios are correct and the incorrect behavior is never exercised. For such use case, pure reasoning model will generally have an advantage as there are not bounded by concrete execution issues.

\begin{lstlisting}[float=htbp, style=mypython, caption={Recursion issue triggered by a non-covered functionality (LiveCodeBench - 2a0f)}, label={listing1}]
class Solution:
    def minimumOperationsToMakeEqual(self, x: int, y: int) -> int:
        if x <= y:
            return y - x
        
        @lru_cache(None)
        def min_operations(n):
            if n <= y:
                return y - n
            return 1 + min(
                n % 11 == 0 and min_operations(n // 11) or float('inf'),
                n % 5 == 0 and min_operations(n // 5) or float('inf'),
                min_operations(n - 1),
                min_operations(n + 1) # Causes infinite recursion
            )
        
        return min_operations(x)
\end{lstlisting}

\subsubsection*{Missing scenario to probe the specification}

We give an example of such case in Listing \ref{listing2}. In that example, the code does not properly account for substring with "s" which starts at another position than 0. Few models run establish such implicit scenario, rather focusing on boundaries for length of "s" or size of "k". Doing so, they occult the bug. When an input from such scenario is generated, models generally identify the error. This signals that, despite prompting for categorical partitioning, this might not be enough to obtain such edge case. Looking at reasoning trace of other approaches, inputs examples are close to never used in the reasoning, showing that it is indeed not a trivial scenario to extract, even more so from the specification only.

\begin{lstlisting}[float=tbp, style=mypython, caption={Missing scenario for substring with non 0 starting position (LiveCodeBench - b863)}, label={listing2}]
class Solution:
    def beautifulSubstrings(self, s: str, k: int) -> int:
        vowels = set("aeiou")
        n = len(s)
        count = 0
        prefix = defaultdict(int)
        prefix[0] = 1
        v_count = 0

        # Should be a second for loop, from i to n
        for i in range(n):
            if s[i] in vowels:
                v_count += 1            
            c_count = (i + 1) - v_count            
            if (v_count * c_count) % k == 0 and (v_count - c_count) % 2 == 0:
                count += prefix[v_count - c_count]            
            prefix[v_count] += 1        
        return count
\end{lstlisting}

\section{Related Works}

\subsection{LLMs for Code, Test and Inputs generations}

The use of large language models (LLMs) for code has been a rapidly growing research area in software engineering. Lemieux et al. proposed CodaMOSA, which combines search-based software testing with Codex to escape coverage plateaus by prompting the model for test cases targeting under-covered functions~\cite{lemieux2023codamosa}. Rao et al. introduced CAT-LM, a GPT-style model pre-trained with an explicit alignment signal between source code and test files, enabling it to generate contextually coherent tests that outperform much larger general-purpose models \cite{rao2023cat}. Chen et al. presented ChatUniTest, a framework that uses an adaptive focal context mechanism and a generate-validate-repair loop to produce compilable and coverage-effective unit tests \cite{chen2024chatunitest}. Dinella et al. propose TOGA \cite{dinella2022toga}, a neural approach that reformulates oracle generation as a ranking problem over a grammar-constrained candidate set, handling both exceptional and assertion oracles. Hossain et al. further improved TOGA by introducing TOGLL \cite{hossain2024togll}, which fine-tunes a suite of code LLMs with targeted prompts. While showing the capacity of LLMs to generate coherent test and codes, recent studies highlight a misalignment between specification, code and tests \cite{huang2024measuring, konstantinou2024llms} which can hamper performance. In particular, findings show that models are prone to asserting the current (potentially buggy) implementation, which limits their fault-detection potential. In contrast, we rely on LLM reasoning using grounded execution via generated test inputs only, limiting the effect of invalid generated oracle. HumanEval+ \cite{humanevalplus} used LLMs along with manual mutator to create additional test inputs to expand a benchmark. Contrary to us, they rely on manual mutator which might not preserve complex object nor constraints of the inputs, leading to additional validation. In particular, we showed that repairing the inputs (especially for atypical datasets such as CoCoClaNeL) has the highest impact on cost of the approach. Direct generation with LLMs allow to leverage reasoning to mitigate this effect.

\subsection{Oracle-Guided Correctness Inference}

Complementary to our work wiht an opposite stance methods, there is a body of literature focusing on oracle-guided correctness inference. Rather than estimating correctness without ground truth, these approaches assume the availability of an existing oracle—typically a reference test suite or a ground-truth solution—and use it to validate, filter, or rank LLM-generated code. The central insight is that passing a sufficiently rich set of pre-existing tests is a reliable proxy for functional correctness, provided the test suite itself is of high quality. AlphaCode \cite{li2022alphacode} operationalises this at scale in competitive programming: it generates millions of candidate programs, filters them against the small set of example test cases provided in the problem statement, and then clusters the surviving candidates before submission. A key limitation of this paradigm is that small test suites systematically overestimate model performance. Mathews et al. \cite{mathews2024tdd} shows that supplying public benchmark test cases directly to the LLM during generation—and iterating on failures—improves results on HumanEval and MBPP, with further gains from a repair loop. Similarly, Fakhoury et al. introduce TiCoder \cite{fakhoury2024ticoder}, a workflow in which a partially formalized intent specification (expressed as user-confirmed test cases) guides iterative code generation. Taken together, oracle-guided approaches differ fundamentally from our method and from oracle-less approaches alike. They require a trusted oracle—either a ground-truth solution, a pre-existing test suite, or a human-confirmed specification—to determine correctness. This makes them highly precise when such an oracle is available, but inapplicable in the typical code evaluation scenario where the oracle is precisely what is missing or in question. Our approach targets this gap: we infer correctness solely from execution traces on generated inputs, without access to ground-truth outputs or any pre-existing test oracle.

\subsection{Oracle-Less Correctness Inference}

The test oracle problem has been recognized as fundamental in software testing. Classic approaches to address it include metamorphic testing, which defines metamorphic relations that relate input/output pairs without requiring explicit oracles, and differential testing, which compares outputs of similar program implementations to detect faults. Yu et al. introduce retromorphic testing, using inverse functions to generate oracles without explicit test cases. Most directly relevant to our work, Valentin et al. propose incoherence, a measure of error estimated without oracles~\cite{valentin2025incoherence}. Fan et al. propose LLMCodeChoice \cite{fan2024oracleguided}, an oracle-guided program selection framework that constructs small distinguishing test suites via differential testing on LLM-sampled programs, using an LLM to predict expected outputs only for those inputs. HoarePrompt~\cite{bouras2025hoareprompt}, which we compare ourselves to, use instead logic based reasoning to infer correctness. Contrary to properties based (metamorphic, intramorphic etc.), we do not require domain knowledge which would require a proper verification of the property (human or other) contrary to code ran inputs. Approaches such as the one from Valentin et al. requires critically multiple code snippets from the same specification (scale issue) and assume incorrect code will differ across each other which might not hold in all cases. In our case, no mutator is needed and only one sample code is needed, the agreement being based solely on inputs. Finally, compared to HoarePrompt we ground the LLMs reasoning via concrete (input, output) execution, which we show help improve the performance and stability of the method.

\section{Threats to Validity}

\textbf{Internal Validity:} Our approach depends on the quality and diversity of generated test inputs. If the categorical partitioning process biases input generation toward certain code patterns, it may not accurately reflect code correctness across diverse specifications, resulting in increased correct code as outlined previously. In practice, however, it offers a better alternative to not generating inputs as we have shown through baselines comparisons. The repair mechanism with a fixed budget may miss valid inputs for complex tasks with unusual requirements or multiple input formats, potentially underestimating the correctness of sophisticated code. This is nonetheless a necessity in order to ensure only valid inputs execute over code. The deduplication step based on code coverage could inadvertently remove important test cases that trigger the same lines but through different logical paths, potentially reducing the effectiveness of outputs verification. This is nonetheless necessary to ensure a minimum number of effective inputs. This issue could be mitigated by adding further information in the deduplication (data-flow process, input size etc.). The binary decision could be affected by LLMs' behavior or non-determinism. In practice, we mitigate this effect by averaging the decision over multiple inputs.

\textbf{External Validity:} Our evaluation is limited to two datasets (LiveCodeBench and CoCoClaNeL). LiveCodeBench is a standardized Python benchmark used in multiple studies. CoCoClaNeL is the unique benchmark HoarePrompt tested on, and so we included it for proper comparison. The generalizability to other code generation domains needs to be explored, yet we believe our approach would translate easily. Indeed, the Code requirements step includes extraction of relevant information (mocks for library, resources needed etc.) which would mean the LLM could realistically generates concrete inputs for more advanced tasks. Our evaluation uses three open-source models of mid-range capability that are fairly recent. While results may not generalize to newer model architectures, frontier closed-source models like GPT-4 or Claude, we chose open-source models among the most recent/capable that could run on a single consumer-grade GPU, making the approach deployable in a lot of situations. Finally, we only evaluated on temperature = $0.5$ only. Results might vary based on this parameters and certain models could improve/decrease this performance based on it. We chose this value following default values used in baselines.

\textbf{Construct Validity:} Our main assumptions for inferring code correctness is that correct code will produce output which the LLMs will recognises as more aligned with the specification on average than for incorrect code. While this assumption is sound, it is impacted by LLM non-determinism. Experiments seem to show the approach is less affected than competing baselines. Moreover, we offset this effect through averaging over multiple inputs. We use MCC and P4 as primary performance metrics. These metrics are standard metrics used in previous studies when considering binary classifier. The threshold selection ($\tau = 0.8$) was chosen empirically based on our experiments. While it seems to hold well across the board, it may not generalize well to new domains or model variants. In that case, leveraging the calibration mechanism we described might be a good alternative. Non-determinism of runs could affect our results. We mitigated this by rerunning each model/dataset/approach three times.

\textbf{Replicability:} We have made sure to describe as thoroughly as possible our steps and processes. All models and datasets used are open-source and available. We also provide a replication package \cite{ReplicationPackage}.

\section{Conclusion}

This paper addresses the critical challenge of validating large language model-generated code by proposing \approach, a novel framework for inferring code correctness through input-grounded reasoning. The fundamental insight of our work is that grounding LLM reasoning with concrete input-output pairs over the specification significantly improves the reliability of correctness assessment. By anchoring reasoning with concrete values rather than allowing models to reason freely over code, we reduce hallucinations and ensure that the specification remains the source of truth throughout the verification process.  While \approach~ incurs higher computational costs than baselines due to input generation and repair, these costs remain comparable to formal verification approaches like HoarePrompt and represent a worthwhile trade-off for improved reliability.

Our comprehensive experimental evaluation demonstrates substantial improvements across multiple dimensions. On two benchmarks with three different LLM architectures, \approach~ achieves up to 39\% improvement in Matthew Correlation Coefficient and 20\% improvement in the P4 metric over Zero-Shot COT reasoning. These gains are made even on CoCoClaNeL, where codes are human written with lower variable semantic, intricate control flow and subtle specifications challenge that pure reasoning approaches might struggle with. Beyond performance, \approach~exhibits superior correct stability: across multiple runs with different random seeds, our approach consistently produces the same verdict more often than baselines, reducing overall false positives and making the approach more suitable for users.

Future research should explore expanding the approach to more complex scenario, optimizing the input generation and repair process through prompt engineering or model fine-tuning or investigating ensemble combinations of \approach~ with other verification approaches to leverage their complementary strengths. We believe \approach~ represents an important step toward making code generation by LLMs a reliable and trustworthy technology for software development, especially for a growing crowd of non-expert user with limited verification skills.

\clearpage
\bibliographystyle{ACM-Reference-Format}
\bibliography{sample-base}

\end{document}